\documentclass[12pt,showkeys, superscriptaddress]{revtex4-2}
\usepackage{graphics,epsfig}
\usepackage{epstopdf}
\usepackage{graphicx}
\usepackage{dcolumn}
\usepackage{amsmath}
\usepackage{epstopdf}
\usepackage{color}
 
\begin{document}
\title{Impact of Perfect Fluid Dark Matter on the Thermodynamics of $AdS$ Ay\'{o}n--Beato--Garc\'{i}a Black Holes}

\author{Amit Kumar}
\email{ammiphy007@gmail.com}
\affiliation{Department of Physics, Institute of Applied Sciences and Humanities, GLA University, Mathura 281406, Uttar Pradesh, India}

\author{Dharm Veer Singh\footnote{Visiting Associate: Inter-University Center of Astronomy and Astrophysics (IUCAA), Pune}}
\email{veerdsingh@gmail.com}
\affiliation{Department of Physics, Institute of Applied Sciences and Humanities, GLA University, Mathura 281406, Uttar Pradesh, India}
 \author{Sudhaker Upadhyay\footnote{Visiting Associate: Inter-University Center of Astronomy and Astrophysics (IUCAA), Pune}}
\email{sudhakerupadhyay@gmail.com}
\affiliation{Department of Physics, K.L.S. College, Nawada, Magadh University, Bodh Gaya, Bihar 805110, India}
\affiliation{Canadian Quantum Research Center 204-3002 32 Ave Vernon, BC V1T 2L7 Canada}
 \affiliation{Department of General \& Theoretical Physics, L. N. Gumilyov Eurasian National University,
Astana, 010008, Kazakhstan}
\begin{abstract}
  \qquad  \qquad \qquad \qquad    \qquad \qquad \qquad\qquad  \textbf{Abstract}\\
 In this paper, we derive the black hole solution in the context of nonlinear electrodynamics (NLED) coupled to a perfect fluid dark matter (PFDM) field. The resulting black hole solution interpolates between the $AdS$ Ay\'{o}n--Beato--Garc\'{i}a (ABG) black hole in the absence of the PFDM field and the Schwarzschild black hole devoid of magnetic monopole charges and PFDM influence. A numerical investigation of the horizon structure and thermodynamic properties, including both local and global stability, is conducted for the obtained black hole solution. The thermodynamic quantities are shown to be modified by the presence of the NLED and PFDM fields. We observe that the behaviour of thermodynamical quantities of black holes depends on these parameters significantly. We also discuss the stability and phase transition dependency on these parameters. 
\end{abstract}
 
\maketitle

\section{\label{sec:level1}Introduction}
A fundamental aspect of general relativity is the identification of exact solutions to Einstein's field equations, a task that often proves to be highly nontrivial. Notable among these solutions are the Schwarzschild black hole, which arises in the vacuum, and the Reissner-Nordstr\"om black hole, which accounts for the presence of charged matter. Both of these solutions exhibit spherical symmetry and possess a central singularity enclosed by an event horizon. However, in contrast to these singular solutions, there also exist non-singular or regular black hole solutions of Einstein's field equations, which avoid the central singularity and represent a distinct class of black hole geometries. Bardeen pioneered the concept of a central matter core based on Sakharov and Gliner proposal \cite{Sakharov:1966,Gliner:1966} and presented a conventional formulation of a black hole with horizons, albeit without a singularity \cite{bar}, 30 years later,    ABG  gave the source of the Bardeen black hole when gravity is coupled to the  NLED \cite{abg}. We can understand the features of NLED   from the perspective of gravity. The concept of NLED was initially put forth by Born and Infeld \cite{born37,born341} to establish the point charge's finite self-energy. Furthermore, significant advancements have been realized in various settings. Since the effective string action naturally yields a generalized Born-Infeld action, NLED also gains significance \cite{lei89,ban08,tak04,hag81, Jafarzade:2020ova}.

The Bardeen black hole represents a spherically symmetric solution that violates the strong energy condition. Bardeen's pioneering work laid the foundation for discovering numerous regular (non-singular) black hole solutions. Among these, a significant example is the $AdS$ ABG black hole, which belongs to this class of regular black hole solutions. The ABG black hole behaves like a de-Sitter core close to its centre ($r\to 0$) and is asymptotically flat, leading to the Reissner-Nordstr\"om black hole in the limit of large $r$ ($r\to \infty$). Consequently, unlike the Reissner–Nordström black hole, the ABG black hole exhibits a de-Sitter core and eventually settles with a regular, non-singular centre. The $AdS$ ABG black hole is a generalized Bardeen solution, as it has been demonstrated to be an exact black hole solution in the $AdS$ spacetime framework \cite{abg11,abg00, Singh:2021nvm}. The effects of magnetic charge  on the phase transition of modified ABG black holes with five parameters
are also studied \cite{jh}.

The thermodynamic properties of asymptotically AdS black holes can be understood as equivalent to thermal states in a conformal field theory (CFT), thanks to the holographic principle. The Hawking-Page transition, a first-order phase transition, holds substantial interest in this context, particularly due to its connection with the thermalization transition in the strongly coupled boundary CFT.
We examine thermodynamic properties and conduct a detailed analysis of its phase structure and critical behaviour of ABG black hole in the presence of a PFDM field. Furthermore, regular black holes have also been found within the context of modified theories of gravity, such as Einstein-Gauss-Bonnet (EGB) gravity \cite{Ghosh:2020tgy, Singh:2019wpu, dvs19, bhu, Myrzakulov:2023rkr}, 4D EGB gravity \cite{Upadhyay:2022axg, EslamPanah:2020hoj, Ghosh:2020ijh, Singh:2020nwo, Singh:2020xju, Paul:2023mlh}, and rotating regular black holes \cite{Ahmed:2022qge, sh11, Singh:2017bwj, rizzo06, Vishvakarma:2023tnl}. By correctly using the NLED Lagrangian parameters, the first law of black hole thermodynamics precisely reproduces the authors' predicted modification to the first law. However, certain regular black hole solutions appear to violate the area law of black hole thermodynamics. This violation stems from inconsistency with the first law of thermodynamics, a fundamental principle that governs the thermodynamic behaviour of black holes. The issue can be addressed by employing a modified form of the first law of thermodynamics, which restores consistency and resolves the apparent discrepancy in the area law for these regular black hole solutions \cite{Maluf:2018lyu, ma14}. Moreover, additional regular solutions with spherical symmetry were proposed \cite{Bronnikov:2000vy, Simpson:2019mud,bambi}. Considerable progress has been made in studying regular black hole solutions along with their properties \cite{fr1, Singh:2017qur,Singh:2020rnm, sabir,Singh:2022xgi, Rodrigues:2022zph,Panotopoulos:2019qjk, Vishvakarma:2023csw,Singh:2022dth, Singh:2022ycn,Singh:2024jgo, Sudhanshu:2024wqb}. 

The structure of this paper is as follows: In Section \ref{sec2}, we explore an action that combines Einstein's gravity with  NLED and a PFDM field, leading to the derivation of a non-singular black hole solution. Section \ref{sec3} is devoted to analysing the black hole's thermodynamic properties, where we find that the solution obeys a modified version of the first law of thermodynamics. Finally, in the concluding section, we summarize the key findings and discuss their broader implications.

\section{An Exact solution of $AdS$ ABG  black hole surrounded by PFDM and NLED fields}\label{sec2}
Let's start with a theory of gravity that includes a PFDM field surrounding an NLED source. The action is described by \cite{Sudhanshu:2024wqb}
\begin{eqnarray}
\mathcal{S} &=&\int d^{4}x\sqrt{-g}\left[  {R}-2\Lambda +\frac{1}{2} \nabla_{a} \varphi\nabla^{a} 
\varphi- V(\varphi) +  {\cal L}_{DM} + {\cal L}(F)\right],
\label{action}
\end{eqnarray}
where ${R}$ is the curvature scalar, and $\Lambda$, representing the cosmological constant, is linked to the $AdS$ length $l$ by $ \Lambda=-{3}/{l^2}$, $\varphi$ is the phantom field, $V(\varphi)$ is phantom 
field potential,  ${\cal L}_{DM}$ denotes the dark matter Lagrangian density, while ${\cal L}(F)$ represents the Lagrangian density of the NLED \cite{abg00}. The explicit expression for ${\cal L}(F)$ is 
written as
\begin{equation}
{\cal L}(F)= \frac{F(1-3\sqrt{2g^2F})}{(1+\sqrt{2g^2F})^{3}}-\frac{3}{2g^2s}\left(\frac{(2g^2F)^{5/4}}{(1+\sqrt{2g^2F})^{5/2}}\right),
\label{nonl1}
\end{equation}
where  $s$ is a free parameter related to the magnetic monopole charge ($g$) and black hole mass $M$ by $s=g/2M$. In the weak field limit, the nonlinear Lagrangian density (\ref{nonl1}) is identified to Maxwell electrodynamics (${{{\cal L}(F)}}=F=F_{ab}F^{ab}/4)$ and also satisfies the weak energy condition ${{{\cal L}(F)}}<0$ and $\partial {{{\cal L}(F)}}/\partial F>0$ \cite{Singh:2022xgi}.
 
The field equations for the metric tensor $g_{ab}$ and the potential $A_a$   corresponding to the action (\ref{action}) are
 \begin{eqnarray}
&&G_{ab} +g_{ab }\Lambda =2\left(\frac{\partial {{{\cal L}(F)}}}{\partial F}F_{ac}F^{c }_b-g_{ab}{\cal L}(F)\right)+ 2\nabla_{a}\varphi\nabla_{b}\varphi
-g_{ab}\nabla_{c}\varphi\nabla^{c}\varphi + T_{ab}^{DM}, \\
  &&\nabla_{a} \left(\frac{\partial {{{\cal L}(F)}}}{\partial F}F^{a b}\right)=0,\qquad\text{and}\qquad \nabla_{a} \left(*F^{ab}\right)=0,
  \label{Field equation}
\end{eqnarray}
where $G_{ab}=R_{ab}-\frac{1}{2}  g_{ab}R$ and $T_{ab}^{DM}$ is an energy-momentum tensor (EMT) corresponding to the dark matter field.
The $T_{ab}^{DM}$ can be approximated as ${\rm diag}(-\rho_{\rm DM}, 0,0,0)$.  Given a particular ansatz, the Maxwell invariant ($F$) and the NLED Lagrangian density   (${\cal L}(F)$) are 
\begin{equation}
    F=\frac{g^2}{2r^4} \qquad \text{and} \qquad {\cal L}(F)=\frac{g^2(r^2-3g^2)}{2(r^2+g^2)^3}+ \frac{8M g^2}{(r^2+g^2)^{5/2}}.
\end{equation}

To solve the field equations, we initially express the general static spherically symmetric metric as \cite{Sudhanshu:2024wqb}
\begin{equation}
ds^2=-f(r)dt^2+h^{-1}(r)dr^2+r^2 \left(d\theta^2+\sin^2\theta d\phi^2\right),\label{met1}
\end{equation}
where $f(r)=e^{\nu(r)}$ and $h(r)=e^{-\mu (r)}$.

 In a static scenario, the component Einstein equations are expressed as
 \begin{eqnarray}
&& G_t^t=e^{-\mu} \left( \frac{1}{r^2} - \frac{\mu^\prime}{r} \right) -\frac{1}{r^2} =  \frac{1}{2} e^{-\mu} \varphi^{\prime 2} -V(\varphi) -\rho_{\rm DM}\,, \label{eq:R-1} \\
&& G_r^r= e^{-\mu} \left( \frac{1}{r^2} + \frac{\nu^\prime}{r} \right) -\frac{1}{r^2} =   -\frac{1}{2} e^{-\mu} \varphi^{\prime 2} -V(\varphi)  \,,  \label{eq:R-2}\\
&& G_{\theta}^{\theta}= \frac{e^{-\mu}}{2} \left( \nu{''} +  \frac{\nu'^2}{2}   + \frac{\nu{'} -\mu{'}}{r} - \frac{\nu^\prime \mu^\prime}{2} \right)  =   \frac{1}{2} e^{-\mu} \varphi^{\prime 2} -V(\varphi). \label{eq:R-3}
\end{eqnarray}
It is intriguing to observe that it is possible to identify solutions that meet the condition ($\mu=-\nu$)  $g_{tt} = -g_{rr}^{-1}$ and the mass density to be $\rho_{\rm DM}=e^{\nu} \varphi^{\prime 2}>0$. Such a solution is possible only due to the signature of the kinetic term of the phantom field, which is in contrast to the ordinary scalar field like the quintessence dark matter, for which no such solution exists.

\begin{figure*}[ht]
\begin{tabular}{c c c c}
\includegraphics[width=.5\linewidth]{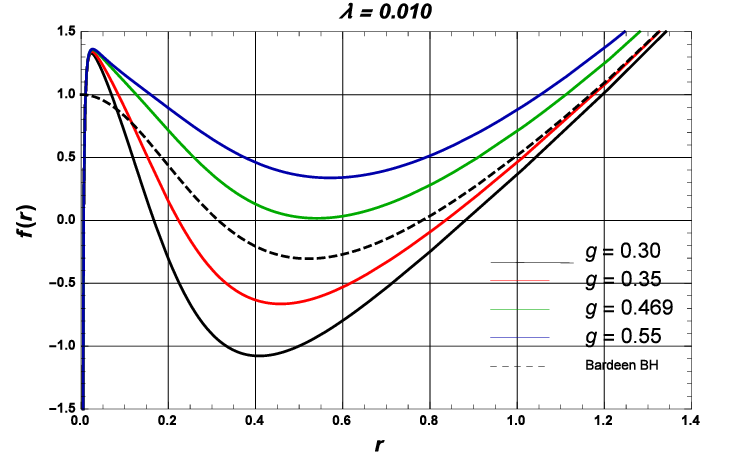}
\includegraphics[width=.5\linewidth]{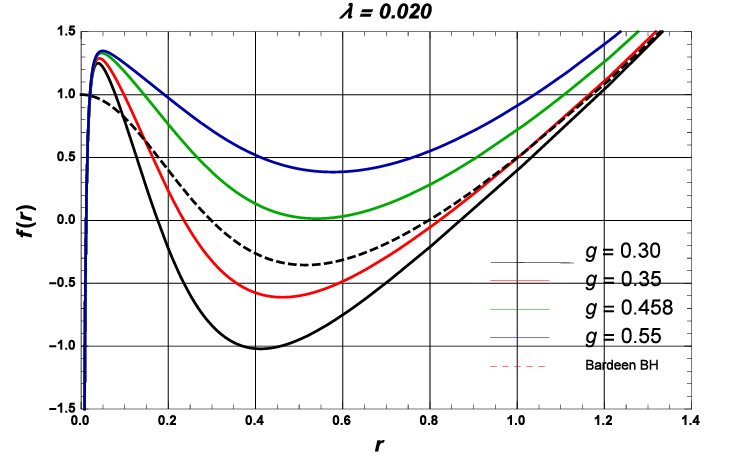}
\end{tabular}
\caption{ $f(r)$ vs $r$  for different values of magnetic monopole charge with a fixed value of scale parameter ($\lambda=0.1 \& \lambda=0.2$).}
\label{fig:h}
\end{figure*}

We set  $\nu= \ln (1-{\cal U)}$ and substitute it into Eqs to find the black hole solution.  (\ref{eq:R-1}) and (\ref{eq:R-3}) and, consequently, we obtain
 \begin{eqnarray}
&& r^2 {\cal U}'' + 2 \epsilon\, r {\cal U}' +2 (\epsilon-1) {\cal U}=\frac{g^2(r^2-3g^2)}{2(r^2+g^2)^3}+ \frac{8M g^2}{(r^2+g^2)^{5/2}},\label{so}
\end{eqnarray}
where $\epsilon$ is constant. The solution to the equation 
 (\ref{so}) is
\begin{eqnarray}
&&{\cal U}=\frac{2M r^2}{(r^2+g^2)^{3/2}}-\frac{g^2r^2}{(r^2+g^2)^2}- \frac{\lambda}{r}   \ln \left( 
\frac{r}{ \lambda } \right)-\frac{r^2}{l^2},  \ \  
  \end{eqnarray}
where  $\lambda$ is a scale parameter, and  the corresponding metric becomes
\begin{eqnarray}
f(r)= 1-\frac{2M r^2}{(r^2+g^2)^{3/2}}+\frac{g^2r^2}{(r^2+g^2)^2}+\frac{\lambda}{r}
\ln\left(\frac{r}{\lambda}\right)+\frac{r^2}{l^2}.
\label{bhs}
\end{eqnarray}
This black hole solution (\ref{bhs}) is characterised by the mass ($M$), magnetic monopole charge ($g$), PFDM field parameter or scale parameter  ($\lambda$) and $AdS$ length related to the cosmological constant via. $(l=\sqrt{-3/\Lambda})$. The solution (\ref{bhs}) interpolates with a PFDM field surrounding the black hole in the limit where the   magnetic monopole charge  is absent 
\begin{eqnarray}
f(r)= 1-\frac{2M }{r}+\frac{\lambda}{r}
\ln\left(\frac{r}{\lambda}\right)+\frac{r^2}{l^2},
\label{bhs}
\end{eqnarray}
and to the  $AdS$ ABG black hole in the limit where the scale parameter ($\lambda$) is absent  \cite{Singh:2021nvm}
\begin{eqnarray}
f(r)= 1-\frac{2M r^2}{(r^2+g^2)^{3/2}}+\frac{g^2r^2}{(r^2+g^2)^2} +\frac{r^2}{l^2}.
\end{eqnarray}
  It is noticed that the obtained black hole solution (\ref{bhs}) behaves with the Reissner–Nordstr\"om black hole asymptotically 
\begin{equation}
f(r)=1-\frac{2M}{r}+\frac{g^2}{r^2}+ \frac{\lambda}{r}   \ln \left( 
\frac{r}{ \lambda } \right)+\frac{r^2}{l^2}+O\left(\frac{1}{r^3}\right).
\end{equation}
The circumstance allows one to estimate the location of the black hole horizon
  \begin{equation}
f(r)=1-\frac{2M r^2}{(r^2+g^2)^{3/2}}+\frac{g^2r^2}{(r^2+g^2)^2}+ \frac{\lambda}{r}   \ln \left( 
\frac{r}{ \lambda } \right)+\frac{r^2}{l^2}=0.
\label{eq.hor}
  \end{equation}
However, Eq. (\ref{eq.hor}) suggests that the black hole horizon's location depends upon the parameters $M,g,\lambda$ and $l$. The Eq. (\ref{eq.hor}) does not have an analytic solution. 
The numerical solution is tabulated in Table \ref{tab:h1}.
\begin{center}
	\begin{table}[h]
		\begin{center}
			\begin{tabular}{l l r l| r l r l r}
\hline
				\multicolumn{1}{c}{ }&\multicolumn{1}{c}{ $\lambda=0.010$  }&\multicolumn{1}{c}{}&\multicolumn{1}{c|}{ \,\,\,\,\,\, }&\multicolumn{1}{c}{ }&\multicolumn{1}{c}{}&\multicolumn{1}{c}{ $\lambda =0.020$ }&\multicolumn{1}{c}{}\,\,\,\,\,\,\\
				\hline
\multicolumn{1}{c}{ \it{$g$}} & \multicolumn{1}{c}{ $r_-$ } & \multicolumn{1}{c}{ $r_+$ }& \multicolumn{1}{c|}{$\delta$}&\multicolumn{1}{c}{$g$}& \multicolumn{1}{c}{$r_-$} &\multicolumn{1}{c}{$r_+$} &\multicolumn{1}{c}{$\delta$}   \\
				\hline
	\,\,\, 0.30\,\,& \,\,0.161\,\, &\,\,  0.885\,\,& \,\,0.724\,\, &\,\, 0.30&0.177\,\,&\,\,0.871\,\,&\,\,0.694\,\,\\
  	\,\, 0.35\,\, & \,\,0.227\,\, &\,\, 0.847\,\,& \,\,0.620\,\,&0.35&\,\, 0.235\,\,&\,\,0.825\,\,&\,\,0.590\,\,
				\\
	\,\,\, 0.469\,\, &  \,\,0.537\,\,  &\,\,0.537\,\,&\,\,0\,\,&0.458\,& \,\, 0.535\,\,&\,\,0.535\,\,&\,\,0\,\,
				\\ 
				\hline
				\hline
    
			\end{tabular}
		\end{center}
		\caption{Inner  horizon ($r_-$), outer horizon ($r_+$) and their difference ($\delta=r_+-r_-$) for different values of magnetic charge $g$ with fixed values of scale parameter ($\lambda= 0.010,  \lambda=0.020$), mass $(M=1)$ and $AdS$ length $(l=1)$.}
				\label{tab:h1}
	\end{table}
\end{center}
The pictorial view of horizon radius is realized for different values of ($g, \lambda$) in Fig (\ref{fig:h}).

The influence of the scale parameter ($\lambda$) and magnetic monopole charge ($g$) on the horizon structure of the derived black hole solution (\ref{bhs}) exhibits the following characteristics:
\begin{enumerate}
    \item The roots of the $f(r)=0$ tells about the number of horizons. The resulting black hole solution possesses three distinct horizons: the Cauchy horizon ($r_-$), the event horizon ($r_+$), and the cosmological horizon ($r_\Lambda$). 
    \item As the magnetic monopole charge ($g$) and the scale parameter ($\lambda$) increase, the black hole's size diminishes correspondingly. 
    \item For ($\lambda=0.010$), the black hole possesses  three horizons when ($g<0.469$) and one horizon when ($g>0.469$). For ($\lambda=0.020$), the black hole has three horizons when ($g<0.458$) and one horizon when ($g>0.458$).
\end{enumerate}

\section{Thermodynamics and Modified First Law}\label{sec3} 
The mass of the black hole solution is determined by applying the condition $f(r)=0$  \cite{Sudhanshu:2024wqb}
\begin{equation}
M_+=    \frac{(r_+^2+g^2)^{3/2}}{2r_+^2}\left(1+\frac{g^2r_+^2}{(r_+^2+g^2)^2}+\frac{r_+^2}{l^2}-\frac{\lambda}{r_+}
\ln\left[\frac{r_+}{\lambda}\right]\right).
\end{equation}
The black hole mass simplifies to that of the $AdS$ black hole in the presence of the PFDM field when the magnetic monopole charge is set to zero, i.e.,
\begin{equation}
M_+=    \frac{r_+}{2}\left(1+\frac{r_+^2}{l^2}-\frac{\lambda}{r_+}
\ln\left[\frac{r_+}{\lambda}\right]\right),
\end{equation}
and the $AdS$ ABG black hole mass in the absence of scale parameter, and the black hole mass interpolates with $AdS$ massive  Reissner-Nordstr\"om black hole mass in the limit $g=\lambda=0$. The conventional method for calculating the Hawking temperature of the black hole is expressed as follows:
\begin{equation}
    T_+=\frac{\kappa}{2\pi}, \qquad \text{where} \qquad \kappa= \frac{\sqrt{-g_{tt}g^{rr}}}{2}.
\end{equation}
In the case of the obtained black hole solution, this leads to the following expression for the temperature:
\begin{eqnarray}
T_+&=&\frac{1}{4\pi r_+}\left[ \frac{2r_+^2}{l^2}-\frac{4g^2r_+^4}{(r_+^2+g^2)^3}+\frac{2g^2r_+^2}{(r_+^2+g^2)^2}-\frac{\lambda}{r_+}\left(1-\ln\left[\frac{r_+}{\lambda}\right]\right)  \right.\nonumber\\
& -&\left. 2\left(1+\frac{r_+^2}{l^2}+\frac{g^2r_+^2}{(r_+^2+g^2)^2}+\frac{\lambda}{r_+}\ln\left[\frac{r_+}{\lambda}\right]\right)\right].
\end{eqnarray}
In the absence of magnetic monopole charge, the temperature of this black hole simplifies to that of the $AdS$ black hole within the context of the PFDM field, represented as follows:
\begin{eqnarray}
T_+=\frac{1}{4\pi r_+}\left[\left(\frac{2r_+^2}{l^2}-\frac{\lambda}{r_+} +\frac{\lambda}{r_+}\ln\left[\frac{r_+}{\lambda}\right]\right)-2\left(1+\frac{r_+^2}{l^2}+\frac{\lambda}{r_+}\ln\left[\frac{r_+}{\lambda}\right]\right)\right].
\end{eqnarray}
The temperature identifies to $AdS$ ABG black hole temperature in the absence of PFDM field, and the black hole temperature interpolates with $AdS$  Reissner-Nordstr\"om black hole temperature in the limit $g=\lambda=0$.
\begin{figure*}[ht]
\begin{tabular}{c c c c}
\includegraphics[width=.5\linewidth]{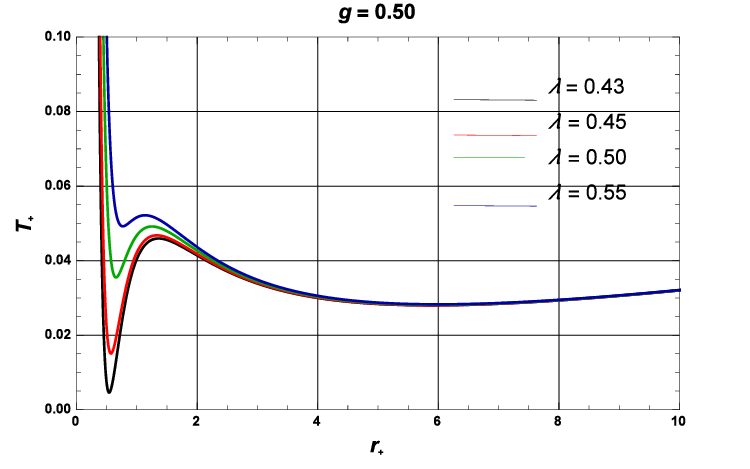}
\includegraphics[width=.5\linewidth]{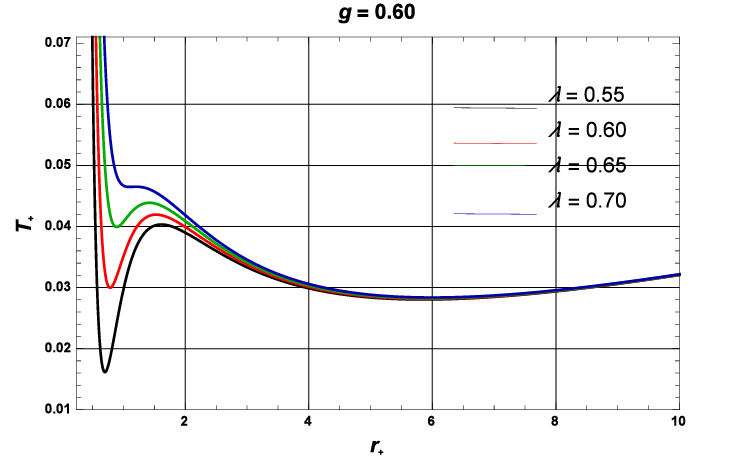}\\
\includegraphics[width=.5\linewidth]{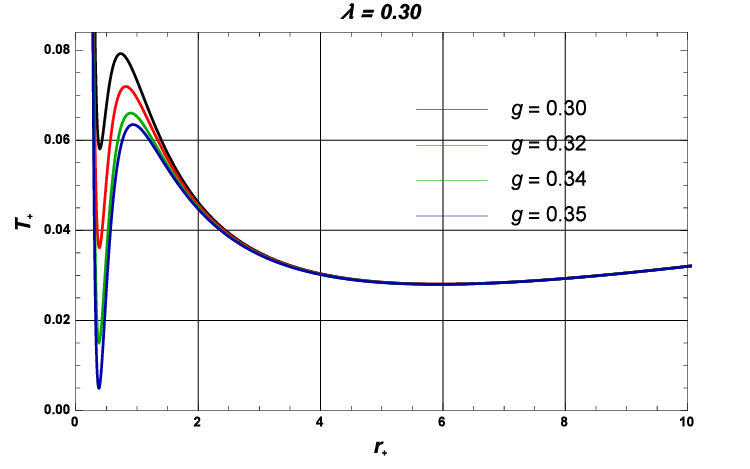}
\includegraphics[width=.5\linewidth]{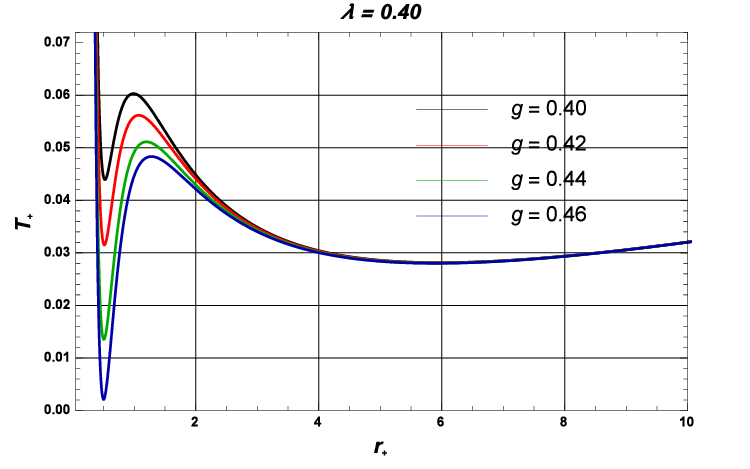}
\end{tabular}
\caption{ The upper panel illustrates the relationship between temperature and horizon radius for various values of the magnetic monopole charge ($g$), with the scale parameter ($\lambda$) held constant. Conversely, the lower panel shows the same relationship for different values of the scale parameter ($\lambda$), while keeping the magnetic monopole charge ($g$) fixed.}
\label{fig:t}
\end{figure*}

 \begin{center}
\begin{table}[h]
\begin{center}
\begin{tabular}{l|l r l r l| r l r r r}
\hline
\multicolumn{1}{c}{ }&\multicolumn{1}{c}{ }&\multicolumn{1}{c}{ }&\multicolumn{1}{c }{$g=0.50$  }&\multicolumn{1}{c}{  }&\multicolumn{1}{c|}{ }&\multicolumn{1}{c}{  }&\multicolumn{1}{c}{ }&\multicolumn{1}{c }{$g=0.60$}&\multicolumn{1}{c }{} &\multicolumn{1}{c }{}\\
\hline
\,\,$\lambda$\,\, &&  \,\, 0.43\,\, &\,\,0.45\,\, &  \,\,0.50\,\, &\,\,0.55\,\,&&\,\,0.55\,\,&\,\,0.60\,\,&\,\,0.65\,\,&\,\,0.70\,\,
\\  \hline
\,\,$T_+^{1} $\,\, &&  \,\, 0.046\,\, &\,\,0.047\,\, &  \,\,0.049\,\, &\,\,0.497\,\,&&\,\,0.040\,\,&\,\,0.042\,\,&\,\,0.044\,\,&\,\,0.046\,\,
\\ 
\,\,$r_{1c}$\,\,&&\,\,1.35\,\, & \,\,1.27\,\,& \,\,  1.25\,\, &\,\,  1.12\,\,&& \,\,1.62\,\,&\,\, 1.51\,\,&\,\,1.44\,\,&\,\,1.37\,\,
\\ 
\,\,$r_{2c} $\,\, &&  \,\, 5.973\,\, &\,\,5.973\,\, &  \,\,5.973\,\, &\,\,5.973\,\,&&\,\,5.995\,\,&\,\,5.995\,\,&\,\,5.995\,\,
\\ 
\,\,$T_+^{2}$\,\,&&\,\,  0.0279\,\, & \,\,0.0279\,\,& \,\,  0.0289\,\, &\,\,  0.0284\,\,&& \,\,0.0282\,\,&\,\, 0.0284\,\,&\,\,0.0286\,\,&\,\,0.0293\,
\\
\hline
\end{tabular}
\end{center}
\caption{The maximum  temperature ($T_+^{Max}$) at critical horizon radius ($r_c $) for varying values of  magnetic monopole charge ($g$) and scale parameter ($\lambda$)  with fixed value of ($l$).}
\label{tab:temp}
\end{table}
\end{center}
Assuming the given black hole solution satisfies the first law of thermodynamics:
\begin{equation}
dM_+=T_+dS_++\phi dg.
\end{equation}
and, utilising the first law of thermodynamics, we can calculate the entropy of the obtained black hole solution.
With the magnetic monopole charge held constant, the first law of thermodynamics provides the following formulation for entropy:
\begin{equation}
S_+=\int \frac{1}{T_+}\frac{dM_+}{dr_+} dr_+.
\label{ent}
\end{equation}
Upon substituting the values of $M_+$ and $T_+$ from the black hole solution (\ref{bhs}) into Eq. (\ref{ent}), the resulting expression for entropy is:
\begin{equation}
S_+= 2\pi r_+\left[\sqrt{r_+^2+g^2}\left(\frac{r_+}{2}-\frac{g^2}{r_+}\right)+\frac{3g^2}{2}\ln\left(r_++\sqrt{g^2+r_+^2}\right)\right].
\end{equation}
It is clear from this expression that the entropy deviates from the area law, implying that the given black hole solution does not adhere to the standard form of the first law of thermodynamics. To address this, we now assume that the black hole obeys the following modified first law of thermodynamics, as proposed in Refs. \cite{Maluf:2018lyu, ma14, dvs19}:
\begin{equation}
 \mathcal{C}(M_+,r_+)\,dM_+=T_+ dS_+,\label{cor}
\end{equation}
The correction factor $\mathcal{C}(M_+,r_+)$ is formulated based on the energy density $T^0_0$ and is expressed as
\begin{equation}
\mathcal{C}(M_+,r_+)=1+4\pi \int_{r_+}^{\infty}r_+^2\frac{\partial T^0_0}{\partial M_+} dr_+.
\end{equation}
In accordance with the modified first law of thermodynamics, the entropy is determined as
\begin{equation}
S_+=\pi r_+^2=\frac{A}{4}.
\label{modent}
\end{equation}
This entropy is consistent with the area law and aligns exactly with the standard entropy of black holes.

We now turn our attention to the stability--both local and global--of this black hole. It is widely recognized that the sign of the heat capacity serves as an indicator of local stability. A positive heat capacity suggests that the system is stable, whereas a negative heat capacity implies instability.
However, Gibbs free energy describes the nature of global stability.
Now, it is a matter of calculation to obtain the heat capacity of the 
obtained black hole solution (\ref{bhs}) from 
the following relation:
\begin{equation}
C_+=\frac{\partial M_+ }{\partial T_+}=\left(\frac{\partial M_+ }
{\partial r_+}\right)\left(\frac{\partial r_+}{\partial 
T_+}\right).
\label{eq:h}
\end{equation}
For the given expressions of  $M_+$  and  $T_+$, the above 
expression (\ref{eq:h}) leads to the heat capacity  as
\begin{equation}
C_+= \frac{2\pi (g^2+r_+^2)^{5/2} 
(3r_+^9+l^2r_+^6(r_++\lambda)g^2(Ag^4+Bg^2+C)-3g^2l^2(g^2+r_+^2))
H}{3r_+^{11}-
l^2r_+^8(r_+-2\lambda)+Dg^8+Eg^6+Fg^4+Gg^2+6g^2l^2(g^2+r_+^2)^2(g
^2+2r_+^2)H},
\label{eq:c}
\end{equation}
where
\begin{eqnarray}
  && A=l^2(\lambda-2r_+), \qquad \qquad  \qquad \qquad B= 3r_+^5-3l^2r_+^2(r_+-\lambda),\\
  && C=6r_+^7+3l^2r_+^4(r_+-3\lambda),\qquad \,\,\qquad D=-l^2(5\lambda-2r_+),\\
  && E=9r_+^5+l^2r_+^2(11r_+-17\lambda),\quad\,\,\, \qquad F=21r_+^7+3l^2r_+^4(r_+-7\lambda),\\
  && G= 15r_+^9+l^2r_+^6(8r_+-11\lambda),\quad\,\,\, \qquad H =\lambda\ln\left[\frac{r_+}{\lambda}\right].
\end{eqnarray}
When the magnetic monopole charge is set to zero, this heat capacity simplifies to that of the $AdS$ black hole surrounded by a PFDM field. In the absence of the PFDM field, it corresponds to the heat capacity of the $AdS$ ABG black hole. Additionally, in the limit where both $g$ and $\lambda$ approach zero, the heat capacity of this black hole interpolates to that of the $AdS$ Reissner-Nordström black hole.
\begin{figure*}[ht]
\begin{tabular}{c c c c}
\includegraphics[width=.5\linewidth]{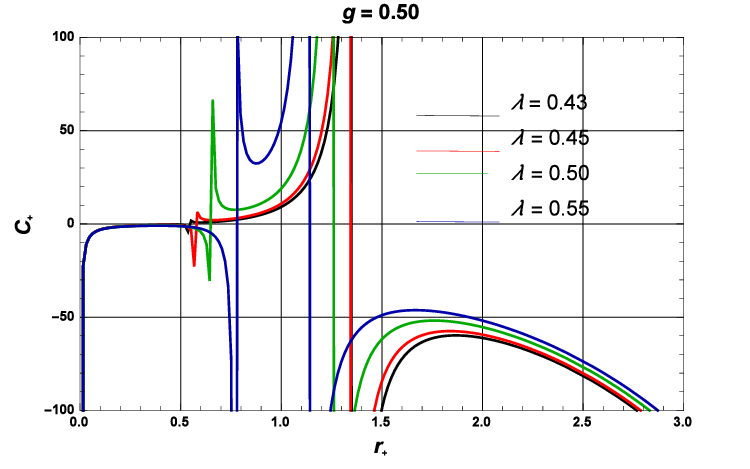}
\includegraphics[width=.5\linewidth]{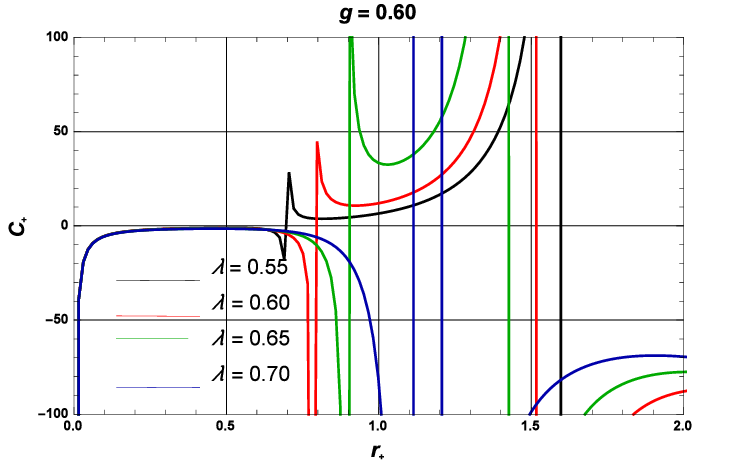}\\
\includegraphics[width=.5\linewidth]{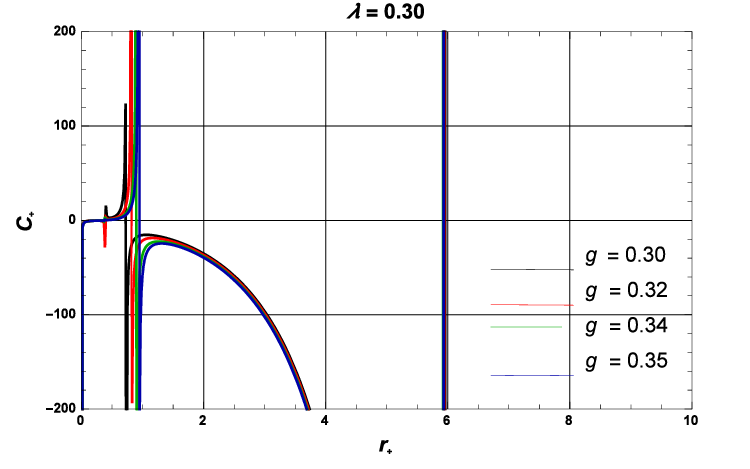}
\includegraphics[width=.5\linewidth]{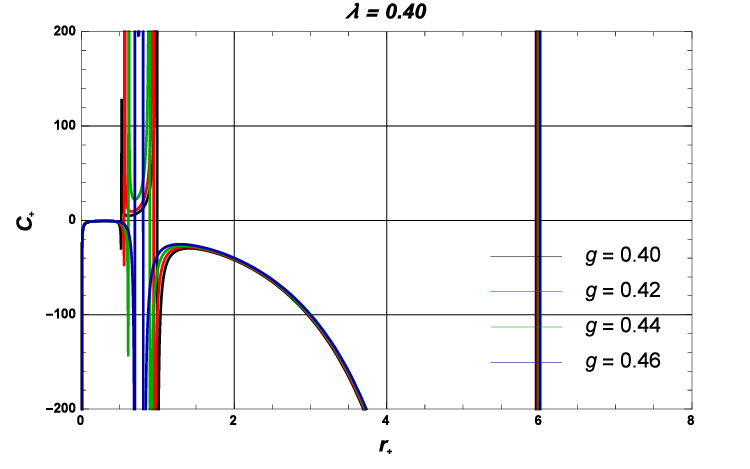}
\end{tabular}
\caption{ The plot of heat capacity vs horizon radius for different values of magnetic monopole charge ($g$) with a fixed value of scale parameter ($\lambda$)(upper panel) and different values scale parameter ($\lambda$) with fixed values of magnetic monopole charge ($g$)  (lower panel).}
\label{fig:c}
\end{figure*}
 
As shown in Fig. \ref{fig:c}, we plotted the expression (\ref{eq:c}) for various scale parameters ($\lambda$) and magnetic monopole charges ($g$) to find out how stable the heat capacity was.
The plot reveals that the heat capacity diverges at the critical points $r_{1+}$ and $r_{2+}$, with $r_{1+} < r_{2+}$ (refer to Fig. \ref{fig:c}). The $AdS$ ABG black hole is stable for horizon radii where $r_+ < r_{1+}$ and $r_+ > r_{2+}$, while it becomes unstable for horizon radii within the range $r_{1+} < r_+ < r_{2+}$. It is evident from Fig. \ref{fig:c} that the $AdS$ ABG black hole experiences two phase transitions: the first occurs at $r_{1+}$, transitioning from a stable state to an unstable one, and the second at $r_{2+}$, where it shifts from an unstable state (when $r_{1+} < r_+ < r_{2+}$) back to a stable state (for $r_+ > r_{2+}$). Specifically, a phase transition takes place at $r_+ = r_{1+} = 0.48$ for $\lambda = 0.1$ and at $r_+ = r_{2+} = 1.3$ for $\lambda = 0.2$, marking the points where the black hole alternates between stable and unstable phases. Moreover, the divergence of the heat capacity at the critical point $r_+ = r_c$ indicates that this represents a second-order phase transition \cite{hp,davis77}. 

The stability of the system can be assessed by examining the heat capacity plots with fixed parameters, as illustrated in Fig. \ref{fig:c}. The heat capacity exhibits discontinuities at two specific points, namely the critical radii $r_+=r_{1+}$ and $r_+=r_{2+}$. At these critical radii, the temperature reaches both its maximum value, denoted as $T_{1+}^{max}$, and its minimum value, referred to as $T_{2+}^{max}$.

Subsequently, we study the black hole's global stability characterized by Gibbs free energy. The Gibbs free energy can be calculated using the definition $G_+=M_+-T_+S_+$, which is expressed as follows: 
\begin{eqnarray}
    G_+&=&  \frac{(r_+^2+g^2)^{3/2}}{2r_+^2}\left(1+\frac{g^2r_+^2}
    {(r_+^2+g^2)^2}+\frac{r_+^2}{l^2}-\frac{\lambda}{r_+}
\ln\left[\frac{r_+}{\lambda}\right]\right)-\frac{r_+}
{4}\left[\left(\frac{2r_+^2}{l^2}-\frac{4g^2r_+^4}
{(r_+^2+g^2)^3}\right.\right.\nonumber\\&+&\left.\left.\frac{2g^2r_+^2}
{(r_+^2+g^2)^2}-\frac{\lambda}{r_+}\left(1-\ln\left[\frac{r_+}
{\lambda}\right]\right)\right)-2\left(1+\frac{r_+^2}{l^2}+\frac{g^2r_+^2}
{(r_+^2+g^2)^2}+\frac{\lambda}{r_+}\ln\left[\frac{r_+}
{\lambda}\right]\right)\right].
\label{eq:g}
\end{eqnarray}
This Gibbs free energy expression simplifies to the free energy of the $AdS$ black hole in the presence of a PFDM field when the magnetic monopole charge is set to zero. In the absence of the scale parameter ($\lambda$), it corresponds to the free energy of the $AdS$ ABG black hole. Furthermore, in the limit where both $g$ and $\lambda$ approach zero, the black hole free energy interpolates with that of the $AdS$ Reissner-Nordstr\"om black hole. The global stability of the black hole is confirmed by the condition $G_+ \leq 0$. We will now examine the behavior of this expression in detail. 
\begin{figure*}[ht]
\begin{tabular}{c c c c}
\includegraphics[width=.5\linewidth]{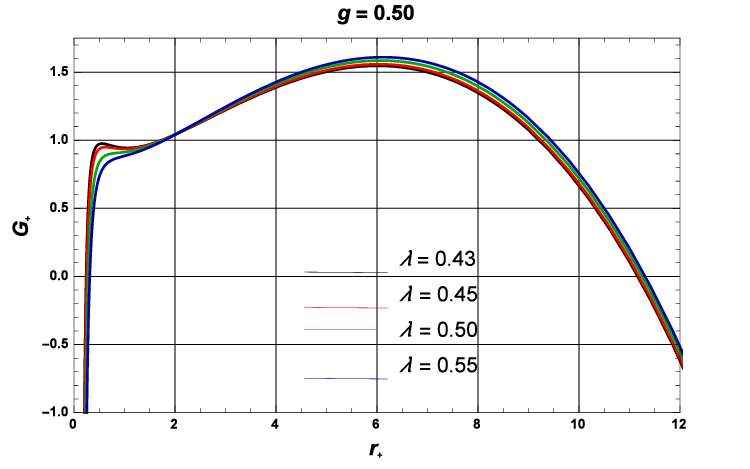}
\includegraphics[width=.5\linewidth]{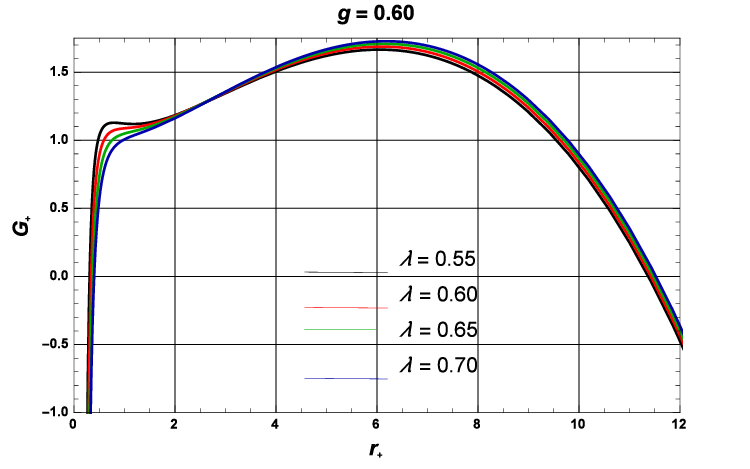}\\
\includegraphics[width=.5\linewidth]{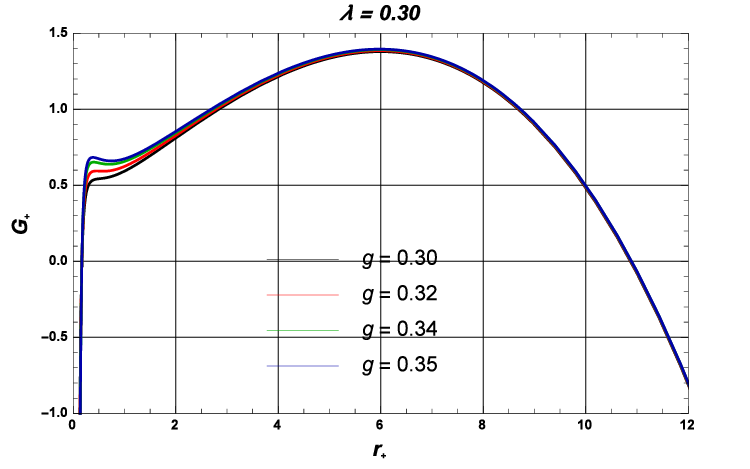}
\includegraphics[width=.5\linewidth]{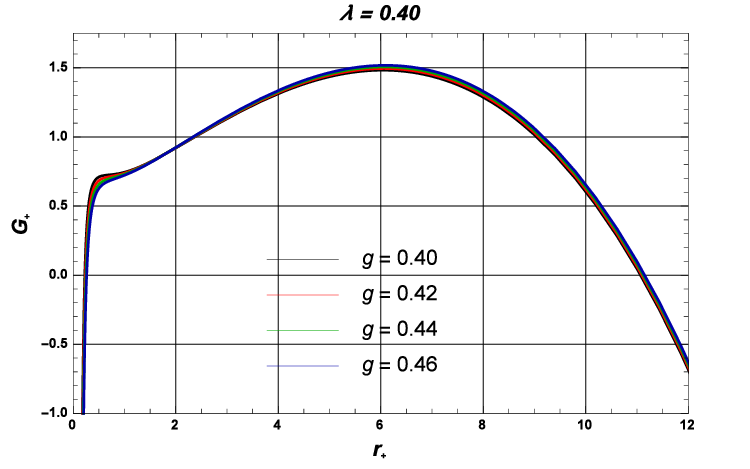}
\end{tabular}
\caption{The upper panel presents a plot of Gibbs free energy as a function of horizon radii for different values of the magnetic monopole charge ($g$), while maintaining a constant scale parameter ($\lambda$). In contrast, the lower panel displays the Gibbs free energy against horizon radii for varying scale parameter values ($\lambda$), with the magnetic monopole charge ($g$) held constant.}
\label{fig:g}
\end{figure*}

In this analysis, we identify a global minimum ($r_{min}$) and a global maximum ($r_{max}$) that correspond to the extremal points of the Hawking temperature. At these critical points, the nature of the free energy undergoes a transformation. Specifically, beyond the minimum radius ${r_{min}}$, the free energy rises with increasing horizon radius $r_+$, reaching its peak at $r_{max}$. Subsequently, beyond this maximum, the free energy begins to decline as the horizon radius increases.

\section{Results and Conclusions}
We have demonstrated that nonsingular black holes coupled with a PFDM field constitute an exact solution within the framework of gravity minimally coupled to NLED. Notably, the well-known Schwarzschild black holes emerge as a special case when both NLED and PFDM fields are absent. 

The characteristics of this solution are defined by a thorough examination of the horizons, which can be maximally three in number: the Cauchy horizon ($r_-$), the event horizon ($r_+$), and the cosmological horizon ($r_\Lambda$). We added a correction that takes regularisation into account so that we could look more closely at the thermodynamic properties and phase transitions of the $AdS$ ABG black holes when the PFDM field is present.

Through this framework, we computed modified thermodynamic quantities, including the Hawking temperature, entropy, and local and global stability measures. Divergences in the heat capacity, which occur at critical locations known as degenerate horizons and where the Hawking temperature reaches a local maximum, are indicators of phase transitions.

The analysis further reveals that stable and unstable branches of the black hole are associated with positive and negative heat capacities, respectively. Intriguingly, the smaller black hole exhibits global stability with negative free energy and positive heat capacity, highlighting its robustness within the thermodynamic landscape.

\section*{Data Availability Statement} 
Data sharing does not apply to this article, as no data sets were generated or analyzed during the current study.

\begin{acknowledgements}  
  DVS thanks to the DST-SERB project (grant no. EEQ/2022/000824).
\end{acknowledgements}


\end{document}